\documentclass[12pt]{amsart}
\usepackage{amsfonts}
\usepackage{amssymb,}
\catcode`\@=11

 \def\AMSTeXfeatures{\Plainheads 
   \let\current@vert=\AMS@vert}

 \def\Plainheads{\sh@ftdiam=0.05em
   \getlabeldims
   \let\vshaftfill=\plnvsolidfill
   \let\hshaftfill=\plnhsolidfill
   \let\th@rhead=\plnrhead
   \let\th@lhead=\plnlhead
   \let\th@dnhead=\plndnhead
   \let\th@uphead=\plnuphead}
 
 \def\glet{\global\let}

 \def\LaTeXfeatures{\catcode`\@=11
   \ifx\@clnwd\undefined \nol@g
      \input ltxcode.tex \dol@g \fi
   \ltxheads \let\current@vert=\new@vert
   \providelto \catcode`\@=\active}

 \def\nol@g{\def\wlog{\edef\garbage}}
 \def\dol@g{\let\wlog=\wl@g} \let\wl@g=\wlog
 \nol@g 

 \newbox\ltobox
 \def\providelto{{\setbox\z@=
   \hbox{$\to$}\minharrlen=\wd\z@
   \global\setbox\ltobox=\hbox{$\activeat>>>$}}
   \def\lto{\mathrel{\copy\ltobox}}}

 \def\ltxheads{\sh@ftdiam=\@wholewidth
   \getlabeldims
   \let\vshaftfill= \ltxvsolidfill
   \let\hshaftfill=\ltxhsolidfill
   \let\th@rhead=\ltxrhead
   \let\th@lhead=\ltxlhead
   \let\th@dnhead=\ltxdnhead
   \let\th@uphead=\ltxuphead}
 {\catcode`\@=\active
   \gdef@#1{\csname #1\string@at\endcsname}
   \glet\activeat=@}
 \def\def@#1{\expandafter\def\csname #1@at\endcsname}

 \def@>#1>#2>{\@rrow R{#1}{#2}}
 \def@<#1<#2<{\@rrow L{#1}{#2}}
 \def@ V#1V#2V{\@rrow V{#1}{#2}}
 \def@ A#1A#2A{\@rrow A{#1}{#2}}
 \def@/#1/#2/#3/{\@rrow{#1}{#2}{#3}}
 \def@.{\ifodd\row\ifmmode\noharrow
     \else\leavevmode.\spacefactor3000 \fi
   \else\novarrow\fi}
 \def@={\ifodd\row\harrow\hequalfill{}{}%
   \else\varrow\vequalfill{}{}\fi}
 \def@:#1{\ifx=#1\harrow\deffill{}{}%
   \else\leavevmode\null:#1\fi}
 \def@|{\current@vert}
  \def\AMS@vert{\varrow\vequalfill{}{}}
  \def\new@vert#1|#2|{\ifodd\row
   \let\nextarrow\vertexvarrow
   \else\let\nextarrow\varrow\fi
   \nextarrow\vshaftfill{#1}{#2}}
 \def@-{\ifmmode\let\next\hl@ne
   \else\let\next\AMSatdash \fi \next}
  \def\hl@ne#1-#2-{\harrow\hshaftfill{#1}{#2}}
  \def\AMSatdash{\let\next\relax\leavevmode
    \def\next@{\ifx\next-%
      \def\next-{\futurelet\next\nextii@}%
     \else\def\next{\hbox{-}}\fi\next}%
    \def\nextii@{\ifx\next-\def\next-{\hbox{---}}%
      \else\def\next{\hbox{--}}\fi\next}%
    \futurelet\next\next@}
 \def@(#1){\tweenarrows{#1}}
 \def@[#1]{\setsp@n#1\relax\activeat}
 \def\fiberbox{\hbox{$\vcenter{\hr@le\hbox{\vr@le
   \kern1ex\vbox{\kern1.2ex}\vr@le}\hr@le}$}}
  \def\hr@le{\hrule height \sh@ftdiam}
  \def\vr@le{\vrule width \sh@ftdiam}
 \def@+#1+#2+#3+{\ifodd\row \harrow{#1}{#2}{#3}%
   \else \varrow{#1}{#2}{#3}\fi}

 \def\Dnarrfill{\vequalfill\Dnhe@d}
 \def\Uparrfill{\Uphe@d\vequalfill}
 
 \def\ontofill{\rtarrfill\kern-0.3em 
   \th@rhead\kern 0.3em} 

 \def\rtarrfill{\hshaftfill\th@rhead}
 \def\ltarrfill{\th@lhead\hshaftfill}
 \def\dnarrfill{\vshaftfill\th@dnhead}
 \def\uparrfill{\th@uphead\vshaftfill}
 \def\hequalfill{\plnhfill=}
 \def\deffill{:\plnhfill=}
 \def\plnvextfill#1{\setbox\z@
   \hbox{\the\textfont3 #1}%
   \dimen@=\dp\z@\advance\dimen@\ht\z@
   \copy\z@ \kern-\dimen@ 
   \cleaders\copy\z@ \vfill
   \kern-\dimen@ 
   \box\z@}
 \def\plnhfill#1{$\m@th\mkern-1.5mu\mathord#1\mkern-6mu
    \cleaders\hbox{$\mkern-2mu\mathord#1\mkern-2mu$}\hfill
    \mkern-6mu\mathord#1\mkern-1.5mu$}
 \def\vequalfill{\plnvextfill{\char'167}}
 \def\plnvsolidfill{\plnvextfill{\char'077}}
 \def\plnhsolidfill{\plnhfill-}
 \def\ltxhsolidfill{\leaders\hrule height\topofshaft depth\botofshaft
   \hfill}
 \def\ltxvsolidfill{\leaders\vrule width\sh@ftdiam\vfill}
 \def\hdashfill{\hd@sh\wd@sh
   \xleaders \hbox{\wd@sh\hd@sh\wd@sh}\hfill
   \wd@sh\hd@sh}
 \def\vdashfill{\vd@sh\wd@sh
   \xleaders \vbox{\wd@sh\vd@sh\wd@sh}\vfill
   \wd@sh\vd@sh}
 \def\dashed{\ifinmeasureCD\else
    \ifodd\row\option{\let\hshaftfill=\hdashfill}%
   \else\option{\let\vshaftfill=\vdashfill}\fi\fi}

 \newdimen\CDstrutht  \newdimen\CDstrutdp
 \newdimen\CDstrutlen \CDstrutlen=\CDstrutht
   \advance\CDstrutlen by \CDstrutdp

 \def\CDstrut{\vrule
   height \ifnum\row=1 \z@\else\CDstrutht \fi
   depth \ifnum\row=\numrows \z@ \else\CDstrutdp \fi
   width\z@}

 \newdimen\CDarrsurr \CDarrsurr=0.375em
 \newdimen\CDdashlen
 \newdimen\CDvarrlen \CDvarrlen=1.5\baselineskip
 \newdimen\minharrlen 
  \setbox\z@\hbox{$\longrightarrow$} \minharrlen=\wd\z@
 \newdimen\minCDharrlen \minCDharrlen=2.5em 
\newdimen \minc@lwd
\def\findminc@lwd{\minc@lwd=2\CDarrsurr
  \advance\minc@lwd\minCDharrlen}

 \newdimen\sh@ftdiam

 \newdimen\labelsurr \labelsurr=1.25 em

\newcount\sp@ncnt \sp@ncnt=\@ne
\newcount\sp@ncnt@ \sp@ncnt@=\@ne
\newdimen\@rrwd \newdimen\@rrdp

 \def\adjustbot#1{\option{\advance\@rrdp#1\relax}}
 
\def\pushvertex#1{\global\p@shlen#1\relax
   \global\let\maybepush=\dopush}

 \newdimen\p@shlen \p@shlen=\z@

 \let\maybepush=\relax
 \def\dopush{\ifinmeasureCD 
   \advance\locdimen by -\p@shlen 
   \else\advance \@rrwd by -\p@shlen \fi 
   \global\let\maybepush=\relax \global\p@shlen=\z@\relax}

 \def\span@ne{\global\sp@ncnt=\@ne\relax}
 \def\setsp@n#1#2{\global\sp@ncnt=#1\relax
   \ifx\relax#2\relax\else\global\sp@ncnt@=#2\relax\fi}

 \def\plnrhead{\llap{$\rightarrow\mkern-1.5mu$}}
 \def\plnlhead{\rlap{$\mkern-1.5mu\leftarrow$}}

 \def\clap#1{\hbox to \z@{\hss #1\hss}}

 \def\plndnhead{\hbox{\the\textfont3 \char'171}}
 \def\plnuphead{\hbox{\the\textfont3 \char'170}}
 \def\Dnhe@d{\hbox{\the\textfont3 \char'177}}
 \def\Uphe@d{\hbox{\the\textfont3 \char'176}}

 \def\ltxrhead{\raise\@xisheight
   \llap{\smash{\@linefnt\@getrarrow(1,0)}}}
 \def\ltxlhead{\raise\@xisheight
   \rlap{\@linefnt\@getlarrow(-1,0)}}
 \def\ltxuphead{\setbox\z@=\rlap{%
   \kern\@halfwidth\@linefnt\char'66}%
   \copy\z@\kern-\ht\z@}
 \def\ltxdnhead{\setbox\z@=\rlap{%
   \kern\@halfwidth\@linefnt\char'77}%
   \ht\z@=\z@\box\z@}

 \def\wd@sh{\kern0.5\CDdashlen}
 \def\hd@sh{\vrule height\topofshaft depth\botofshaft
    width\CDdashlen}
 \def\vd@sh{\hrule height\CDdashlen
   depth\z@ width\sh@ftdiam}

\def\xylist{14{3434}13{2414}12{1723}%
  23{1413}34{1153}11{0867}43{0707}%
  32{0580}21{0414}31{0291}41{0}}
\newcount\tgtcnt@
\def\find@xyargs{\dimen@=\@rrdp
  \advance\dimen@ by \CDstrutlen
  \tgtcnt@=\dimen@ \dimen@=\@rrwd 
  \divide\dimen@ by \@m 
  \divide \tgtcnt@ by \dimen@ 
  \expandafter\testxy\xylist\relax
  \unitlength=\@xarg\@rrdp
  \divide\unitlength by\@yarg\relax}
\def\testxy#1#2#3{\ifnum\tgtcnt@>#3
    \@xarg=#1\relax \@yarg=#2\relax
    \let\next=\ignorerest
  \else\let\next\testxy\fi\next}
\def\ignorerest#1\relax{\relax}

\let\scalefactor=\@ne
\def\SWarrow{\find@xyargs\vector
  (-\@xarg,-\@yarg)\scalefactor\hskip-\wd\@linechar}
\def\NWarrow{\find@xyargs\vector
  (-\@xarg,\@yarg)\scalefactor\hskip-\wd\@linechar}
\def\NEarrow{\find@xyargs\vector
  (\@xarg,\@yarg)\scalefactor}
\def\SEarrow{\find@xyargs\vector
  (\@xarg,-\@yarg)\scalefactor}
\def\rightupline{\find@xyargs\@linelen=\scalefactor
     \unitlength\@sline}
\def\rightdownline{\find@xyargs\@yarg=-\@yarg\relax
     \@linelen=\scalefactor\unitlength\@sline}

\def\Sim{\ifodd\row\setbox\z@=\hbox{$\sim$}\dimen@=\ht\z@
 \advance\dimen@ by -\@xisheight
  \vbox{\box\z@\kern-\@xisheight\kern\dimen@}%
  \else\hbox{$\wr$}\fi}

\def\harrow#1#2#3{\inmeasureCDtrue\findminarrwd
  {#2}{#3}{\sp@ncnt\minharrlen}\inmeasureCDfalse\span@ne
  \mathrel{\hbox{\options\hplace{#1}\ulabel{#2}\dlabel{#3}}}}

\def\noharrow{\harrow\hfill{}{}}
\def\vertexvarrow#1#2#3{\findarrdp \@rrwd=\z@ \setsp@n\@ne\@ne
  \vbox to \z@{\kern-1.2\CDstrutht
  \rlap{\options\vplace{#1}\llabel{#2}\rlabel{#3}}\vss}}

\newif\ifinmeasureCD
\def\measurelabel#1{\setbox\z@
  \hbox{$\scriptstyle#1\kern\labelsurr$}%
  \ifdim\wd\z@>\@rrwd \@rrwd=\wd\z@\fi}
\def\findminarrwd#1#2#3{\@rrwd=#3\relax
   \measurelabel{#1}\measurelabel{#2}}
\def\findCDarrwd#1#2{\@rrwd=\minCDharrlen
   \measurelabel{#1}\measurelabel{#2}%
  }

\newcount\row \row=\@ne \newcount\col \col=\@ne 
 \newcount\numrows 
\numrows=\@ne
 \newcount\numcols
\newcount\arrspan \newdimen\vrtxhalfwd  \newbox\tempbox

\def\DANABUG{\advance\col by \@ne
 \@rrwd=\minCDharrlen
  \advance\@rrwd by \vrtxhalfwd
  \advance\@rrwd by \CDarrsurr
  \ifnum\col>\numcols \numcols=\col
     \newlocdimen{col\the\col}\locdimen=\@rrwd 
  \else \ifdim\@rrwd>\c@l \c@l=\@rrwd\fi\fi}

\def\drop#1\\{
  \findvrtxhalfsum\DANABUG\advance\row by 2 \measureinit}

\def\measureinit{\col=\@ne \vrtxhalfwd=-\CDarrsurr\arrspan=\@ne\@rrwd=\z@
   \setbox\tempbox=\hbox\bgroup$}
\def\measure{
  \let\harrow\measureCDarrow
  \let\CDCR=\measureCR 
   \findminc@lwd 
  \inmeasureCDtrue
  \row=\@ne \numcols=\z@ \measureinit}

\def\endmeasure{\findvrtxhalfsum\DANABUG
  \numrows=\row 
  \inmeasureCDfalse}

\def\newlocdimen#1{\advance\dimenc@unt by \@ne
  \ifnum\dimenc@unt<\insc@unt
     \else\errmessage{No room for the CD}\fi
  \dimendef\locdimen=\dimenc@unt
  \expandafter\dimendef\csname#1\endcsname=\dimenc@unt}

 \def\r@wc@l{\csname row\the\row col\the\col\endcsname}
 \def\c@l{\csname col\the\col\endcsname}

 \def\findvrtxhalfsum{$\egroup
  \newlocdimen{row\the\row col\the\col}
  \locdimen=\vrtxhalfwd 
  \vrtxhalfwd=0.5\wd\tempbox 
  \advance\vrtxhalfwd by \CDarrsurr
  \advance\locdimen by \vrtxhalfwd 
  \advance\@rrwd by \locdimen 
  \maybepush
  \divide\@rrwd by \arrspan\relax
  \ifdim\@rrwd<\minc@lwd
    \ifnum\col>\@ne \@rrwd=\minc@lwd\fi \fi
  \loop 
    \ifnum\col>\numcols \numcols=\col
       \newlocdimen{col\the\col}
       \locdimen=\@rrwd 
    \else \ifdim\@rrwd>\c@l \c@l=\@rrwd\fi \fi
   \ifnum\arrspan>\@ne
      \advance\arrspan by -1 \advance\col by \@ne
  \repeat }

 \def\measureCDarrow#1#2#3{\findvrtxhalfsum
   \arrspan=\sp@ncnt\relax\global\sp@ncnt=1\relax
   \advance\col by \@ne
   \findCDarrwd{#2}{#3}%
   \setbox\tempbox=\hbox\bgroup$}

 \newcount\dr@tn \dr@tn=\z@
 \def\locate#1:#2{\ifinmeasureCD\else
   \count@=-#1
   \multiply\count@ by 2
   \advance\count@ by #2
   \dimen@=\count@\@rrwd
   \ifnum\dr@tn=\@ne\relax \else\dimen@=-\dimen@ \fi
   \dimen@i=\@rrdp
   \ifnum\dr@tn>\z@\advance\dimen@i by \CDstrutlen \fi
   \dimen@i=\count@\dimen@i
   \count@=#2 \multiply\count@ by 2
   \divide\dimen@ by \count@
   \divide\dimen@i by \count@
   \lift\dimen@i\nudge\dimen@\fi}

\def\betweenCDrows{\advance\row by \@ne \col=\@ne
\options}

\def\hbegin{\hbox\bgroup\kern\c@l \kern-\r@wc@l$}
\def\hend{$\glet\maybepush\relax \CDstrut\egroup}
\def\vbegin{\setbox\tempbox=\hbox\bgroup$}
\def\vend{$\egroup\ht\tempbox=\z@\dp\tempbox\CDvarrlen
  \box\tempbox}
\def\setCD{\let\harrow=\setCDarrow
  \let\CDCR=\setCR 
  \row=\@ne \col=\@ne \hbegin}
\let\endsetCD=\hend 

\def\findarrwd{\@rrwd=\z@ \count@=\col \advance\count@ by\sp@ncnt
  \loop\ifnum\count@>\col \advance\count@ by -1
      \advance\@rrwd by\csname col\the\count@\endcsname\repeat}
\def\setCDarrow#1#2#3{\kern\CDarrsurr\advance\col by \@ne
  \findarrwd \advance\@rrwd by -\r@wc@l  
  \@rrdp=\z@ 
  \maybepush
  \advance\col by -\@ne \advance\col by \sp@ncnt \span@ne
  \hbox to \@rrwd{\options
   \@rrwd=\scalefactor\@rrwd\hss
   \hplace{#1}\ulabel{#2}\dlabel{#3}\hss}%
   \kern\CDarrsurr}

\newdimen\labspacei 
\newdimen\labspaceii 

\newdimen\@xisheight
  \@xisheight=\the\fontdimen22\textfont2
\newdimen\labelskip
  \labelskip=\the\fontdimen10\textfont3 
\newdimen\topofshaft
\newdimen\botofshaft
\newdimen\botofulabel
\newdimen\topofdlabel
\def\getlabeldims{
  \topofshaft=0.5\sh@ftdiam
  \botofshaft=\topofshaft
  \advance\topofshaft by \@xisheight  
  \advance\botofshaft by -\@xisheight  
  \botofulabel=\topofshaft
  \advance\botofulabel by \labelskip
  \topofdlabel=\botofshaft
  \advance\topofdlabel by \labelskip}

\def\ulabel{\ifnum\row=\@ne\let\next\ulabeli
   \else\let\next\ulabellap\fi\next}
\def\ulabeli#1{\vbox{
  \clap{\kern-\@rrwd$\scriptstyle#1$}%
  \kern\botofulabel}\maybeoffset}
\def\ulabellap#1{\vbox to \z@{\vss
  \clap{\kern-\@rrwd$\scriptstyle#1$}%
  \kern\botofulabel}\maybeoffset}
\def\dlabel{\ifnum\row=\numrows\let\next\dlabeli
   \else\let\next\dlabellap\fi\next}
\def\dlabeli#1{\vtop{\kern\topofdlabel
  \clap{\kern-\@rrwd$\scriptstyle#1$}%
  }\maybeoffset}
\def\dlabellap#1{\vbox to \z@{\kern\topofdlabel
  \clap{\kern-\@rrwd$\scriptstyle#1$}%
  \vss}\maybeoffset}
\def\rlabel#1{\vbox to \z@{\vss
  \rlap{\kern\labelskip$\scriptstyle#1$}%
  \vss\kern-\@rrdp}\maybeoffset}
\def\llabel#1{\vbox to \z@{\vss
  \llap{$\scriptstyle#1$\kern\labelskip}%
  \vss\kern-\@rrdp}\maybeoffset}
\def\swlabel#1{\vtop{\kern0.5\@rrdp
  \llap{$\scriptstyle#1$\kern\labelskip\kern-0.5\@rrwd}
  }\maybeoffset}
\def\nwlabel#1{\vbox{
  \llap{$\scriptstyle#1$\kern\labelskip\kern-0.5\@rrwd}%
  \kern-0.5\@rrdp}\maybeoffset}
\def\selabel#1{\vtop{\kern0.5\@rrdp
  \rlap{\kern0.5\@rrwd\kern\labelskip$\scriptstyle#1$}%
  }\maybeoffset}
\def\nelabel#1{\vbox{
  \rlap{\kern0.5\@rrwd\kern\labelskip$\scriptstyle#1$}%
  \kern-0.5\@rrdp}\maybeoffset}
\def\cplace#1{\vbox to \z@{\vss
  \clap{$#1$\kern-\@rrwd}%
  \kern-\@rrdp\vss}\maybeoffset}
\def\hplace#1{\hbox to \@rrwd{#1}\maybeoffset}
\def\vplace#1{\clap{\vbox to \z@{#1\kern-\@rrdp}}\maybeoffset}

\newdimen\nudgeamount \nudgeamount=\z@
\newdimen\liftamount \liftamount=\z@
\let\maybeoffset\relax
\newbox\offsetbox \newdimen\lastheight
\def\dooffset{
  \setbox\offsetbox=\lastbox \lastheight=\ht\offsetbox 
  \setbox\offsetbox=\vbox{\kern-\liftamount\box\offsetbox}%
  \ht\offsetbox=\lastheight
  \kern\nudgeamount\box\offsetbox\kern-\nudgeamount
  \global\nudgeamount=\z@ \global\liftamount=\z@
  \glet\maybeoffset=\relax}
\def\nudge#1{\ifinmeasureCD\else
  \global\advance\nudgeamount#1\relax
  \global\let\maybeoffset\dooffset\fi}
\def\lift#1{\ifinmeasureCD\else
  \global\advance\liftamount#1\relax
  \global\let\maybeoffset\dooffset\fi}

\def\findarrdp{\@rrdp=\CDvarrlen
  \ifnum\sp@ncnt@>1
    \advance\@rrdp by \CDstrutlen
    \multiply\@rrdp by \sp@ncnt@
    \advance\@rrdp by -\CDstrutlen \fi
 }

\def\varrow#1#2#3{\ifnum\sp@ncnt>\@ne 
     \sp@ncnt@=\sp@ncnt\relax\fi
  \findarrdp \@rrwd=\z@ 
  \kern\c@l
   \hbox to \z@{\options
   \@rrdp=\scalefactor\@rrdp
    \hss\vplace{#1}\llabel{#2}\rlabel{#3}\hss}%
  \global\advance\col by \@ne \setsp@n\@ne\@ne
  }

\def\novarrow{\varrow\vfill{}{}}

\def\tweenarrows#1{\findarrwd \findarrdp \setsp@n\@ne\@ne
  \rlap{\options\cplace{#1}}}

\def\usarrow #1#2#3{\dr@tn=\@ne
  \findarrwd \findarrdp \setsp@n\@ne\@ne 
  \rlap{\options\cplace{#1}\nwlabel{#2}\selabel{#3}}%
  \dr@tn=\z@}
\def\dsarrow #1#2#3{\dr@tn=\tw@
  \findarrwd \findarrdp \setsp@n\@ne\@ne 
  \rlap{\options\cplace{#1}\swlabel{#2}\nelabel{#3}}%
  \dr@tn=\z@}
 \def\@rrow#1{\csname #1@rrow\endcsname}
 \def\R@rrow{\harrow \rtarrfill}
 \def\L@rrow{\harrow \ltarrfill}
 \def\V@rrow{\varrow \dnarrfill}
 \def\A@rrow{\varrow \uparrfill}
 \def\SE@rrow{\dsarrow \SEarrow}
 \def\NW@rrow{\dsarrow \NWarrow}
 \def\SW@rrow{\usarrow \SWarrow}
 \def\NE@rrow{\usarrow \NEarrow}
 \def\DS@rrow{\dsarrow \dnslope}
 \def\US@rrow{\usarrow \upslope}
 \def\upslope{\find@xyargs
       \@linelen=\unitlength\@sline}
 \def\dnslope{\find@xyargs\@yarg=-\@yarg\relax
       \@linelen=\unitlength\@sline}

\newtoks\optionlist 
\optionlist={}
\let\options\relax
\def\dooptions{\the\optionlist\global\optionlist={}%
  \glet\options=\relax}
\def\option#1{\ifinmeasureCD\else
  \glet\options=\dooptions
  \global\optionlist=\expandafter{\the\optionlist\relax#1}\fi}
\def\wider#1{\ifinmeasureCD\else
   \option{\advance\@rrwd by #1}\fi}
\def\deeper#1{\ifinmeasureCD\else
   \option{\advance\@rrdp by #1}\fi}

{\def\\{\global\let\sptoken= }\\ }

\def\CR{\futurelet\nexttok\testCR}
\def\testCR{\ifx\nexttok\sptoken
   \let\next\eatspaceCR\else\let\next\CDCR\fi\next}
\def\eatspaceCR#1 {\CR}
\def\measureCR{\ifx\nexttok\endmeasure\let\nextCR\relax
    \else\let\nextCR\drop\fi\nextCR}
\def\setCR{\ifodd\row
  \ifx\nexttok\endsetCD\else\hend\betweenCDrows\vbegin\fi
  \else\vend\betweenCDrows\hbegin\fi}

\countdef\dimenc@unt=11
\def\CD#1\endCD{
   \begingroup\let\\=\CR
  \m@th\offinterlineskip
   \measure#1\endmeasure\null\,\vcenter{\setCD#1\endsetCD}\,
   \endgroup
    }

\ifx\@clnwd\undefined \nol@g\else\catcode`\ =14\relax\fi
 \font\@linefnt=line10 
 \newcount\@tempcnta
 \newcount\@tempcntb
 \newdimen\@tempdima
 \newdimen\@tempdimb
 \newdimen\@wholewidth
 \newdimen\@halfwidth
   \@wholewidth\fontdimen8\@linefnt \@halfwidth .5\@wholewidth
 \newdimen\unitlength
 \newcount\@xarg
 \newcount\@yarg
 \newcount\@yyarg
 \newbox\@linechar
 \newdimen\@linelen
 \newdimen\@clnwd
 \newdimen\@clnht
 \newif\if@negarg
 
 \def\@whilenoop#1{}

 \def\@whiledim#1\do #2{\ifdim #1\relax#2\@iwhiledim{#1\relax#2}\fi}

 \def\@iwhiledim#1{\ifdim #1\let\@nextwhile=\@iwhiledim 
         \else\let\@nextwhile=\@whilenoop\fi\@nextwhile{#1}}

 \def\@sline{\ifnum\@xarg< 0 \@negargtrue \@xarg -\@xarg \@yyarg -\@yarg
   \else \@negargfalse \@yyarg \@yarg \fi
 \ifnum \@yyarg >0 \@tempcnta\@yyarg \else \@tempcnta -\@yyarg \fi
 \ifnum\@tempcnta>6 \@badlinearg\@tempcnta0 \fi
 \ifnum\@xarg>6 \@badlinearg\@xarg 1 \fi
 \setbox\@linechar\hbox{\@linefnt\@getlinechar(\@xarg,\@yyarg)}%
 \ifnum \@yarg >0 \let\@upordown\raise \@clnht\z@
    \else\let\@upordown\lower \@clnht \ht\@linechar\fi
 \@clnwd=\wd\@linechar
 \if@negarg \hskip -\wd\@linechar \def\@tempa{\hskip -2\wd\@linechar}\else
      \let\@tempa\relax \fi
 \@whiledim \@clnwd <\@linelen \do
   {\@upordown\@clnht\copy\@linechar
    \@tempa
    \advance\@clnht \ht\@linechar
    \advance\@clnwd \wd\@linechar}%
 \advance\@clnht -\ht\@linechar
 \advance\@clnwd -\wd\@linechar
 \@tempdima\@linelen\advance\@tempdima -\@clnwd
 \@tempdimb\@tempdima\advance\@tempdimb -\wd\@linechar
 \if@negarg \hskip -\@tempdimb \else \hskip \@tempdimb \fi
 \multiply\@tempdima \@m
 \@tempcnta \@tempdima \@tempdima \wd\@linechar \divide\@tempcnta \@tempdima
 \@tempdima \ht\@linechar \multiply\@tempdima \@tempcnta
 \divide\@tempdima \@m
 \advance\@clnht \@tempdima
 \ifdim \@linelen <\wd\@linechar
    \hskip \wd\@linechar
   \else\@upordown\@clnht\copy\@linechar\fi}
 
 \def\@getlinechar(#1,#2){\@tempcnta#1\relax\multiply\@tempcnta 8
 \advance\@tempcnta -9 \ifnum #2>0 \advance\@tempcnta #2\relax\else
 \advance\@tempcnta -#2\relax\advance\@tempcnta 64 \fi
 \char\@tempcnta}
 
 \def\vector(#1,#2)#3{\@xarg #1\relax \@yarg #2\relax
 \@tempcnta \ifnum\@xarg<0 -\@xarg\else\@xarg\fi
 \ifnum\@tempcnta<5\relax
 \@linelen=#3\unitlength
 \ifnum\@xarg =0 \@vvector 
   \else \ifnum\@yarg =0 \@hvector \else \@svector\fi
 \fi
 \else\@badlinearg\fi}
 
 \def\@svector{\@sline
 \@tempcnta\@yarg \ifnum\@tempcnta <0 \@tempcnta=-\@tempcnta\fi
 \ifnum\@tempcnta <5
   \hskip -\wd\@linechar
   \@upordown\@clnht \hbox{\@linefnt  \if@negarg 
   \@getlarrow(\@xarg,\@yyarg) \else \@getrarrow(\@xarg,\@yyarg) \fi}%
 \else\@badlinearg\fi}
 
 \def\@getlarrow(#1,#2){\ifnum #2 =\z@ \@tempcnta='33\else
 \@tempcnta=#1\relax\multiply\@tempcnta \sixt@@n \advance\@tempcnta
 -9 \@tempcntb=#2\relax\multiply\@tempcntb \tw@
 \ifnum \@tempcntb >0 \advance\@tempcnta \@tempcntb\relax
 \else\advance\@tempcnta -\@tempcntb\advance\@tempcnta 64
 \fi\fi\char\@tempcnta}
 
 \def\@getrarrow(#1,#2){\@tempcntb=#2\relax
 \ifnum\@tempcntb < 0 \@tempcntb=-\@tempcntb\relax\fi
 \ifcase \@tempcntb\relax \@tempcnta='55 \or 
 \ifnum #1<3 \@tempcnta=#1\relax\multiply\@tempcnta
 24 \advance\@tempcnta -6 \else \ifnum #1=3 \@tempcnta=49
 \else\@tempcnta=58 \fi\fi\or 
 \ifnum #1<3 \@tempcnta=#1\relax\multiply\@tempcnta
 24 \advance\@tempcnta -3 \else \@tempcnta=51\fi\or 
 \@tempcnta=#1\relax\multiply\@tempcnta
 \sixt@@n \advance\@tempcnta -\tw@ \else
 \@tempcnta=#1\relax\multiply\@tempcnta
 \sixt@@n \advance\@tempcnta 7 \fi\ifnum #2<0 \advance\@tempcnta 64 \fi
 \char\@tempcnta}
\catcode`\ =10

\dol@g 
\catcode`\@=\active
\LaTeXfeatures

\begin{document}

\def\al{Val}
\numberwithin{equation}{section}

\newtheorem{theorem}{Theorem}[section]
\newtheorem{lemma}[theorem]{Lemma}

\newtheorem*{theorema}{Theorem A}
\newtheorem*{theorema1}{Theorem A${}^\prime$}
\newtheorem*{theoremb}{Theorem B}
\newtheorem*{corc}{Corollary C}

\newtheorem{prop}[theorem]{Proposition}
\newtheorem{proposition}[theorem]{Proposition}
\newtheorem{corollary}[theorem]{Corollary}
\newtheorem{corol}[theorem]{Corollary}
\newtheorem{conj}[theorem]{Conjecture}
\newtheorem{sublemma}[theorem]{Sublemma}
\newtheorem{quest}[theorem]{Question}

\theoremstyle{definition}
\newtheorem{defn}[theorem]{Definition}
\newtheorem{example}[theorem]{Example}
\newtheorem{examples}[theorem]{Examples}
\newtheorem{remarks}[theorem]{Remarks}
\newtheorem{remark}[theorem]{Remark}
\newtheorem{algorithm}[theorem]{Algorithm}
\newtheorem{question}[theorem]{Question}
\newtheorem{subsec}[theorem]{}
\newtheorem{clai}[theorem]{Claim}
\newtheorem{problem}{Problem}

\renewcommand*{\theproblem}{\arabic{problem}}

\def\toeq{{\stackrel{\sim}{\longrightarrow}}}
\def\into{{\hookrightarrow}}

\def\wt{\widetilde}

\def\kp{Val}


%
\title[Automata and automata mappings of semigroups]
 {Automata and automata mappings of semigroups}

\author[Boris Plotkin]{{\bf B.~Plotkin, }\ \  {\bf T.~Plotkin}\\
\\
 Institute of Mathematics \\
 Hebrew University, 9190401 Jerusalem, Israel\\}

\author[ Tatjana Plotkin]{\\
 Department of Mathematics \\
 Bar-Ilan University, 5290002 Ramat Gan, Israel\\
 plotkin at macs.biu.ac.il}


\maketitle

\begin{abstract}
The paper is devoted to two types of algebraic models of automata. The usual (first type) model leads to the
developed decomposition theory (Krohn-Rhodes theory). We introduce another type of automata model and
study how these automata are related to cascade connections of automata of the first type. The introduced
automata play a significant role in group theory and, hopefully, in the theory of formal languages.
\end{abstract}

\medskip



\section{Introduction}\label{sec:intro}

In this small note we consider two types of algebraic model of automata. In Section \ref{sec:first_type_atm} we consider usual automata, i.e., triples of the form $(A, X, B)$ where $A$ is a set of states of the automaton, $X$ a set of inputs, and a set $B$ is treated as the set of external states. The theory of such automata is well-known (see for example \cite{E}, \cite{KRT}, \cite{PGG}) and leads to the famous Krohn-Rhodes decomposition theory.  In Section \ref{sec:second_type_atm} we introduce the notion of automata of the second type.  These are triples of the form $(A, \Gamma, \Sigma)$ where $A$ is a set of states of the automaton, $\Gamma$ a semigroup of inputs, and
  a semigroup $Y$ is treated as  outputs.  The triple $(A, \Gamma, \Sigma)$ is provided by two binary operations subject to special conditions. This notion is motivated by the ideas of the paper \cite{GNS}.  One of the aims of what follows and of entire note is to show how these automata are related to usual ones, and how they appear in the process of the cascade connections of the automata of the first type. Sections \ref{sec:CC_1_type} and \ref{sec:CC_2_type} deal with this relation.

\section{The first type of automata}\label{sec:first_type_atm}
Let $(A,X,B)$ be a triple with two operations $\circ$ and $\ast$, such that $a \circ x \in A$, $a \ast x = b \in B$ for $a \in A$, $b \in B$.
Here $A$ is a set of states, $B$ is a set of external states and $X$ is a system of inputs. Such a triple is said to be a {\emph {
pure automaton of the first type}}.

We define a {\emph{semigroup automaton of the first type}} as a triple $(A, \Gamma, B)$ with  the semigroup of inputs $\Gamma$, and operations $\circ: A\times X\to A$ and $\ast: A\times X\to B$,  subject to conditions
 $$a \circ \gamma_1 \gamma_2 = (a \circ \gamma_1) \circ \gamma_2,$$ 
 $$a \ast \gamma_1 \gamma_2 = (a \circ \gamma_1) \ast \gamma_2,$$

 \noindent where $\gamma_1 , \gamma_2 \in \Gamma$.
There are no operations on the set $B$. These are the usual definitions of  pure and semigroup automata (see \cite{PGG}).


Given sets $A$ and $B$, denote by $S_A$ the semigroup of transformations of the set $A$ and by $Fun(A,B)$ the set of mappings from $A$ to $B$. Consider Cartesian product $S_{A,B} = S_A \times Fun(A,B)$. Here $S_{A,B}$ is a semigroup with respect to the multiplication: $(\sigma_1, \varphi_1)(\sigma_2, \varphi_2)= (\sigma_1 \sigma_2, \sigma_1 \varphi_2)$, $\sigma \in S_A$, $\varphi \in Fun(A,B)$. Define an automaton $(A,S_{A,B},B)$ by the rule: $a \circ (\sigma, \varphi) = a \sigma$, $a \ast (\sigma, \varphi) = a \varphi$. Every automaton $(A,X,B)$  
is determined by a mapping $X \to  S_{A,B}$.

The automaton $(A,S_{A,B},B)$ is a semigroup automaton. Any semigroup automaton $(A, \Gamma, B)$ is determined by a homomorphism $\Gamma \to S_{A,B}$. In this sense the automaton $(A,S_{A,B},B)$ is universal.

Let again $(A,X,B)$ be a pure automaton. We have a mapping $X \to S_{A,B}$. Let $F(X)$ be the free semigroup over the set $X$. The initial mapping is extended up to a homomorphism $F(X) \to S_{A,B}$, which determines a semigroup automaton $(A, F(X), B)$. We can pass from $(A, F(X), B)$ to a faithful semigroup automaton $(A,\Gamma,B)$ where $\Gamma$ is a result of factorization of the semigroup $F(X)$ by the kernel of homomorphism in $ S_{A,B}$. So, any pure automaton $(A,X,B)$ gives rise to a faithful semigroup automaton $(A,\Gamma,B)$. This allows to construct a decomposition theory  for pure automata (Krohn-Rhodes theory)\cite{KRT}, \cite{PGG}.

\section{The second type of automata}\label{sec:second_type_atm}

We define a {\emph{pure automaton of the second type}} as a triple $(A, X, Y)$, where $A$ is a set of states, $X$ a system of inputs and $Y$ a system of outputs. This triple is equipped with operations $\circ: A\times X\to A$ and $\ast: A\times X\to Y$.

The axiom for the operation $
\ast $ for the second type automaton is different from that of the first type.
The difference comes up from the fact that in the second case the set of outputs $Y$ is intended to be used  as input signals  in the serial connection of two automata of the first type, and thus should satisfy the conditions below.

Define {\emph{ a semigroup automaton of the second type}} as a triple  $(A, \Gamma, \Sigma)$  with the set of states $A$, semigroup of inputs $\Gamma$, semigroup of outputs $\Sigma$ and  operations $\circ: A\times \Gamma\to A$ and $\ast: A\times \Gamma\to \Sigma$,  subject to conditions
$$
a \circ \gamma_1 \gamma_2 = (a \circ \gamma_1) \circ \gamma_2,
$$
$$
a \ast \gamma_1 \gamma_2 = (a \ast \gamma_1) ((a \circ \gamma_1) \ast \gamma_2).
$$
 Let us study how an arbitrary pure automaton of the second type  $(A, X, Y)$ gives rise to a semigroup automaton of the second type $(A, \Gamma, \Sigma)$.


Consider first the situation when $\Gamma = F(X)$ and $\Sigma =F(Y)$, the free semigroups.
The transition $ a \to a \circ x$ determines the mapping $X \to S_A$ and then the homomorphism $F(X) \to S_A$. We have $a \circ u \in A$ and thus $a \circ u_1 u_2 = (a \circ u_1) \circ u_2$.

Proceed now from the mapping $\alpha : A \times F(X) \to F(Y)$ with the condition $\alpha (a, u_1 u_2) = \alpha (a, u_1) \ \alpha(a \circ u_1, u_2)$. This condition arises from the definition of the cascade connection of automata of the first type.

Denote $\alpha(a,u)$ by $a\ast u$. Then $a \ast u_1 u_2 =(a \ast u_1) ((a \circ u_1) \ast u_2)$. We got a semigroup automaton of the second type $(A, F(X), F(Y))$. As a rule, this kind of automata with free semigroups  $ F(X)$, $F(Y)$ is used in applications (as in \cite{GNS}). We are interested here in a more general case of arbitrary semigroups  $\Gamma$, $\Sigma$.


Suppose the  automaton $(A, F(X), F(Y))$ is given and we need to define $(A, \Gamma, \Sigma)$. Proceed from surjections (homomorphisms) $\mu : F(X) \to \Gamma$ and $\nu: F(Y) \to \Sigma$ and point out conditions which lead to the automaton $(A, \Gamma, \Sigma)$.  Define $a \circ u = a \circ u ^ \mu$, $a \in A$, $ u \in F(X)$. Then  the semigroup $\Gamma$ acts in $A$. Define the relation between $\mu$ and $\nu$ as $$a \ast u^\mu = (a\ast u)^\nu.$$ Then we calculate $$a \ast (u_1 u_2)^\mu = a \ast u_1 ^\mu \ u_2^\mu = (a \ast u_1 u_2)^\nu = ((a \ast u_1)((a\circ u_1)*u_2))^\nu =$$
$$= (a \ast u_1)^\nu ((a \circ u_1) \ast u_2)^\nu = (a \ast u_1 ^\mu)((a \circ u_1 ^ \mu) \ast u_2^\mu).$$ So, $a \ast u_1^\mu u_2^\mu = (a \ast u_1 ^\mu)((a \circ u_1 ^ \mu) \ast u_2^\mu),$ as required.
Thus, the defined above relation between $\mu$ and $\nu$ allows us to construct the semigroup automaton $(A, \Gamma, \Sigma)$, grounding on a pure automaton of the second type $(A,X,Y)$.
Various other automata $(A, \Gamma, \Sigma)$ can be constructed using  cascade connections of the automata of the first type (see Section \ref{sec:CC_2_type}).

\section{Cascade connections of automata of the first type }\label{sec:CC_1_type}

We start the topic of constructions in automata theory. Cascade connections discussed here generalize parallel and serial connections of automata.

Let (pure) automata $(A_1, X_1, B_1)$ and $(A_2, X_2, B_2)$ be given. Their cascade connection has the form $(A_1 \times A_2, X, B_1\times B_2)$. In order to make this triple an automaton it is assumed that  the mappings
$$\alpha : A_2  \times X \to X_1, \ \ \beta : X \to X_2.$$
\noindent are defined. We set:
$$(a_1, a_2) \circ x = (a_1 \circ \alpha(a_2,x), a_2\circ \beta(x)),$$
$$(a_1, a_2) \ast x = (a_1 \ast \alpha(a_2,x), a_2\ast \beta(x)).$$

Hence, the cascade connection $(A_1 \times A_2, X, B_1\times B_2)$ is determined by a triple $(X, \alpha, \beta)$. Define further a category of such triples. Let $\mu:(X, \alpha, \beta) \to (X', \alpha ', \beta ')$ be a morphism. We have commutative diagrams of mappings

$$
\CD
X\times A_2 @>\alpha >> X_1\\
@. @/SE/\mu// @AA\alpha' A\\
@. X'\times A_2\\
\endCD
\qquad
\CD
X @>\beta >>\ X_2\\
@. @/SE/\mu// @AA\beta' A\\
@.\ X',\\
\endCD
$$

\noindent
 Here, $\mu(x,a_2) = (\mu(x), a_2)$. The category of triples determines the category of cascade connections of the given automata.

Proceed to semigroup  automata. Given $(A_1, \Gamma_1, B_1)$ and $(A_2, \Gamma_2, B_2)$, pass to $(A_1 \times A_2, \Gamma, B_1\times B_2)$. We need here a triple $(\Gamma, \alpha, \beta)$, where the mapping $\beta: \Gamma \to \Gamma_2$ is a homomorphism of semigroups and the mapping $\alpha: A_2\times\Gamma  \to \Gamma_1$ satisfies condition similar to those for a homomorphism, namely, $\alpha (a_2, \gamma_1 \gamma _2)= \alpha(a_2,\gamma_1 )\alpha(a_2 \circ \beta(\gamma_1),\gamma_2 )$.

We define actions $\circ$ and $\ast$ in a cascade connection of the  automata $(A_1, \Gamma_1, B_1)$ and $(A_2, \Gamma_2, B_2)$ as above, and obtain a category of triples $(\Gamma, \alpha, \beta)$. It is checked that an automaton $(A_1 \times A_2, \Gamma, B_1\times B_2)$ satisfies the axioms of a semigroup automata, and we have a category of such automata. It is proved that such a category has the universal terminal object, called  wreath product of the given automata and denoted by
$$(A_1, \Gamma_1, B_1)\ wr \ (A_2, \Gamma_2, B_2).$$
By the definition of a terminal object, every cascade connection of the given automata is embedded into wreath product.


Recall that the terminal object is realized as follows \cite{PGG}. We consider  a triple $(\Gamma, \alpha, \beta)$, where $\Gamma$ is a wreath product of semigroups
$$\Gamma=\Gamma_1 \ wr^{A_2} \ \Gamma_2.$$
It is defined as follows: take the semigroup $\Gamma_1^{A_2}$ whose elements are mappings $\bar \gamma_1 : A_2 \to \Gamma_1$, $\bar \gamma_1 (a_2) = \gamma_1 \in  \Gamma_1$. The semigroup $\Gamma_2$ acts in $\Gamma_1 ^{A_2}$ by the rule $(\bar \gamma_1 \circ \gamma_2)(a_2) = \bar \gamma_1 (a_2 \circ \gamma_2)$.
Let $\Gamma$ be the Cartesian product $\Gamma = \Gamma_1^{A_2} \times \Gamma_2$ with the multiplication  defined by the rule $(\bar \gamma_1 , \gamma_2)(\bar \gamma_1 ' , \gamma_2 ')= (\bar \gamma_1 \cdot (\bar \gamma_1 ' \circ \gamma_2), \gamma_2 \gamma_2 ')$. 
 Then $\Gamma_1 \ wr^{A_2} \ \Gamma_2$. Setting $\alpha( a_2,(\bar \gamma_1, \gamma_2)) = \bar \gamma_1 (a_2)$ we define $\alpha : A_2\times \Gamma  \to \Gamma_1$. Setting $\beta(\bar \gamma_1, \gamma_2)= \gamma_2$ we get $\beta:\Gamma \to \Gamma_2$. The necessary conditions are checked and we come to the automaton
$$(A_1 \times A_2, \Gamma_1 wr^{A_2} \Gamma_2, B_1 \times B_2) = (A_1, \Gamma_1, B_1) \ wr \ (A_2, \Gamma_2, B_2).$$

The wreath product construction works in the Krohn-Rhodes theory which leads to the decomposition of pure automata and to the definition of complexity of this decomposition.

\section{Automata of the second type and cascade connections} \label{sec:CC_2_type}

Now we want to relate the automata of the second type with the cascade connection operation defined in the previous section. Let us show that any automaton of the second type   can be built  through the serial connection of automata.

  Recall that a serial connection of automata $(A_1, \Gamma_1, B_1)$ and $(A_2, \Gamma_2, B_2)$  is a particular case of the cascade connection and defined by the triple $(\Gamma, \alpha, \beta)$ where $\Gamma=\Gamma_2$,
  the mapping $\beta: \Gamma \to \Gamma_2$ is defined as $\beta(x)=x$   and the mapping $\alpha:  A_2 \times \Gamma_2\to \Gamma_1$ satisfies  $\alpha (a_2, \gamma_1 \gamma _2)= \alpha(a_2, \gamma_1)\alpha(a_2 \circ \gamma_1,\gamma_2 )$.

 It is enough two consider a serial connection of semiautomata of the form  $(A,\Gamma)$ and $(B,\Sigma)$. Given a
 map $\alpha:A\times\Gamma\to\Sigma$ with the condition
$$
 \alpha(a,\gamma_1\gamma_2)=\alpha(a,\gamma_1)\alpha((a\circ \gamma_1),\gamma_2),
 $$
define $ \alpha(a,\gamma)=a\ast\gamma$. Then
$$
 a \ast\gamma_1\gamma_2=(a\ast\gamma_1)((a\circ \gamma_1)\ast\gamma_2).
 $$
Here the map $\bar a:\Gamma\to\Sigma$ is defined by $\bar a(\gamma)=a\ast\gamma$ for each $a\in A$. This map is not correlated with the multiplications in $\Gamma$ and $\Sigma$. We call such a mapping an {\emph{ automaton one}}. It is not a homomorphism of semigroups but something similar. Automata mappings of semigroups  are determined by  automata of the second type.

Action of the semigroup $\Gamma$ in $A\times B$ is defined by the rule
$$
(a,b)\circ\gamma=((a\circ\gamma),(b\circ(a\ast\gamma))).
$$
So we have a triple $(A,\Gamma,\Sigma)$ with the action of $\Gamma$ in $A$ defined by $a\circ\gamma$ and with $a\ast \gamma \in\Sigma$. 
This is the automaton of the second type corresponding to the serial connection  $(A,\Gamma)$ and $(B,\Sigma)$.

Since we proceed from the automaton $(A,X,Y)$, the condition $\bar a (x) = y \in Y$ holds for every $x \in X$. Note that if $\bar a : X \to Y$ is a bijection here, then $\bar a : F(X) \to F(Y)$ is a bijection as well.
We use the fact that if $u = v$ is an equality in the free semigroup $F(Y)$, then it is an identity there, and $u$ and $v$ coincide graphically. Induction by the length of the word finishes the proof. In particular, for the words $wx$ and $w' \ x'$ we have $a \ast wx = (a \ast w)((a \circ w) \ast x) = (a \ast w) y$ and $a \ast w' \ x' =(a \ast w')y'$. If $a \ast wx = a \ast w' \ x'$, then $y=y'$ and $a \ast w = a \ast w'$, which leads to the uniqueness.

\section{Applications}\label{sec:appl}

 Let, for example, $\Gamma$ and $\Sigma$ coincide with the free semigroup $F(X)$ and the mapping $\bar a: F(X) \to F(X)$ is bijective. Take a group generated by all such mappings $\bar a$. For the finite sets $A$ and $X$ this group is called automaton group.
 Numerous applications of automata groups in group theory are described in the seminal paper by Grigorchuk-Nekrashevich-Sushchanskii (\cite{GNS}). This group is used also in the theory of formal languages, since there are some transformations of words which are elements of $F(X)$ (see \cite{TT}).
\begin{remark}
In the paper \cite{GNS} also the case of linear automata with acting free semigroup is treated. For this case one can introduce the notion of a linear automaton of the second type. Replacing serial connection of pure automata by the triangular product of linear automata (see \cite{PGG}) the result similar to that of Section \ref{sec:CC_2_type} can be obtained. Of course, the axioms for the linear automaton of the second type will be different from the axioms for the pure case.
\end{remark}


\begin{thebibliography}{BGK2}

\bibitem {E} Eilenberg, S.:  Automata, languages and machines.  Academic Press (1976)
\bibitem {GNS} Grigorchuk, R., V.Nekrashevich, V.,  Sushchanskii, V,.: Automata, dynamical systems and infinite groups, Proc. Steklov Inst. Math. v.231 (2000), 134-214
\bibitem {KRT} Krohn, K., Rhodes, J., Tilson, B.: Lectures on finite semigroups. in: Algebraic theory
of machines, languages and semigroups, Academic Press, NY (1968)
\bibitem {PGG}Plotkin, B., Greenglaz,L., Gvaramija,A.: Algebraic structures in automata and databases theory. World Scientific Publ., Singapore - New-Jersey (1992)
\bibitem {TT} Tesson, P., Therien, D., : Logic meets algebra: the case of regular languages, Logical Methods in Computer Science
Vol. 3 (1:4) 2007, pp. 1–37

\end{thebibliography}
\end{document}